\documentstyle[aps,epsfig,prb,floats]{revtex}
\tightenlines

\begin{document}
\twocolumn[
\hsize\textwidth\columnwidth\hsize\csname@twocolumnfalse\endcsname
\draft
\title
{Raman scattering through surfaces having biaxial symmetry
}

\author{
A. Gozar$^{1,2}$
}

\address{
$^{1}$Bell Laboratories, Lucent Technologies, Murray Hill, NJ 07974 \\
$^{2}$University of Illinois at Urbana-Champaign, Urbana, IL
61801-3080\\
}
\date{\today}
\maketitle

\begin{abstract}

Magnetic Raman scattering in two-leg spin ladder materials and the relationship between the anisotropic exchange integrals are analyzed by P. J. Freitas and R. R. P. Singh in Phys. Rev. B, {\bf 62}, 14113 (2000).
The angular dependence of the two-magnon scattering is shown to provide information for the magnetic anisotropy in the Sr$_{14}$Cu$_{24}$O$_{41}$ and La$_{6}$Ca$_{8}$Cu$_{24}$O$_{41}$ compounds. 
We point out that the experimental results of polarized Raman measurements at arbitrary angles with respect to the crystal axes have to be corrected for the light ellipticity induced inside the optically anisotropic crystals.
We refer quantitatively to the case of Sr$_{14}$Cu$_{24}$O$_{41}$ and briefly discuss potential implications for spectroscopic studies in other materials with strong anisotropy. 

\end{abstract}

\bigskip

]
\narrowtext

In a recent article (see Ref.~\cite{Freitas}) the authors developed an elegant theory on the two-magnon Raman scattering in two-leg ladder systems.
Within the Fleury-Loudon model \cite{Fleury} the interaction of light with the magnetic degrees of freedom can be written as:
\begin{equation}
H_{int} = \sum_{<i,j>} J'_{ij} ({\bf e}_{in} \cdot \hat{r}_{ij}) ({\bf e}_{out} \cdot \hat{r}_{ij}) {\bf S}_{i} \cdot {\bf S}_{j} ,
\label{eq1}
\end{equation}
where ${\bf e}_{in}$, ${\bf e}_{out}$ are the polarization versors of the incoming and outgoing field, $\hat{r}_{ij}$ are unit vectors along the bond directions, and $J'_{ij}$ represent the Raman coupling constants between spins $i$ and $j$.
The key result of Ref.~\cite{Freitas} is the expression for the magnetic scattering cross section in parallel polarization:
\begin{equation}
I_{||}(\omega, \theta) = I_{||}(\omega, 0) [\cos^2(\theta) -
\frac{J'_{||}}{J'_{\perp}} \frac{J_{\perp}}{J_{||}}  \sin^2(\theta)]^{2}
\label{eq2}
\end{equation}
Here $I(\omega, \theta)$ is the scattering intensity at angle $\theta$ with the rung direction and $J_{||,\perp}$ are the magnetic exchange integrals along and across the ladder legs.
Equation (\ref{eq2}) would allow in principle the determination of the magnetic exchanges $J_{||,\perp}$ by \emph{continuous} variation of the angle $\theta$ if the $J'_{||} / J'_{\perp}$ ratio is known \cite{Freitas} as destructive interference cancels the intensity at the critical angle $\theta_{cr}$~=~$\arctan [(J'_{\perp} J_{||} / J'_{||} J_{\perp})^{1/2}]$.

We want to stress that although relation (\ref{eq2}) is based on \emph{correct} theoretical assumptions, from the experimental point of view it cannot be applied in a straightforward way for $\theta \neq m \pi / 2$ where $m$ is an integer.
The theoretical prediction for the angular dependence is contrasted with the experimental results in Fig.~1, which shows the dependence of the two-magnon scattering intensity in the ladder compound Sr$_{14}$Cu$_{24}$O$_{41}$ \cite{Gozar1} in parallel polarization as a function of the angle between the light polarization and the rungs direction.
The intensity minimum does not reach zero but it remains 50\% of the maximum value.

\begin{figure}[t]
\centerline{
\epsfig{figure=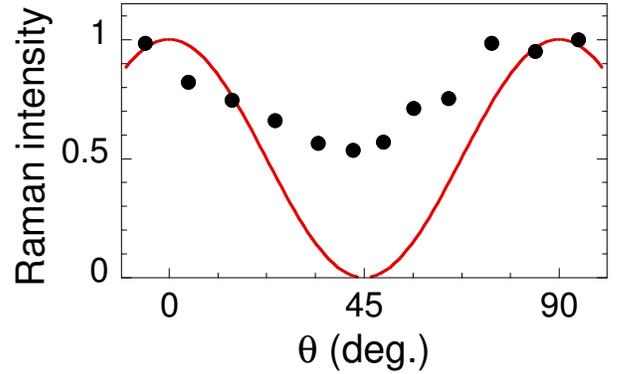,width=80mm}
}
\caption{
Two-magnon peak intensity in parallel polarization (dots) as a function of the angle $\theta$ between the light polarization and the rungs direction for Sr$_{14}$Cu$_{24}$O$_{41}$.
The solid line is the theoretical curve predicted from Eq.~(\ref{eq2}) for $ J'_{||}J_{\perp} / J'_{\perp}J_{||} = 1 $
}
\label{Fig.1}
\end{figure}

It is known that anisotropic media allow in principle two plane waves with two different phase velocities and two different polarizations to propagate in any given direction \cite{Born}.
When $\theta \neq m \pi / 2$, the projections of the electric field on the crystal axes will propagate with different phase velocities and an elliptical polarization is induced as a function of the penetration depth.
In two-leg ladders this effect can be far from negligible in the visible spectrum usually used in Raman spectroscopy. 
In Sr$_{14}$Cu$_{24}$O$_{41}$ for example, for a spectrum taken with incident excitation energy $\omega_{i} = 1.837$~eV, numerical values for the real and imaginary parts of the complex refraction index $\tilde{n} (\omega_{i}) = n (\omega_{i}) + i k (\omega_{i})$ extracted from Kramers-Kronig analysis of the reflectivity data, \cite{Motoyama} give $|n_{x} - n_{y}| \approx 0.376$, $k_{x} = 0.0945$ and $k_{y} = 0.282$ where $x, y$ denote the rung and leg directions, respectively.
On the characteristic penetration depth scales $\tilde{\lambda}_{x,y}$ the phase differences between the components of the electric field $E_{x,y}$ given by $\delta(z) = (2 \pi / \lambda_{i}) (n_{x} - n_{y})z $ are significant, namely $\delta(\tilde{\lambda}_{x,y}) = 0.63 \pi$ and $0.21 \pi$, respectively.
For the particular Raman geometry used to derive Eq.~(\ref{eq2}) we have to use in Eq.~(\ref{eq1}) for the inelastically scattered photons at $z$ inside the sample the formulas:
\begin{equation}
{\bf e}_{in,out} (z, \omega) = \frac{\cos(\theta) e^{-\alpha_{x}z} \hat{x} + \sin(\theta) e^{-\alpha_{y}z} e^{\pm i \delta(z)} \hat{y}}{(\cos^2(\theta) e^{-2\alpha_{x}z} + \sin^2(\theta) e^{-2\alpha_{y}z})^{1/2}}
\label{eq3}
\end{equation}
where $\alpha_{x,y}$ are absorbtion coefficients and the $+$ and $-$ correspond to the incoming and outgoing electromagnetic field.
The polarization versors become dependent on the penetration depth $z$ and on the energy of the incoming and scattered light $via$ the parameter $\delta$ and the absorbtion coefficients $\alpha_{x,y}$.
Furthermore, the determination of the total scattered intensity requires an integration over the $z$ coordinate. 
A real comparison with the measured signal is quantitatively difficult not only because the calculations are laborious but also because the propagation of any experimental errors in the determination of the optical parameters is hard to control.
There are also other optical corrections to be taken into account - transmission at the sample interface, absorbtion and reabsorbtion processes of the incoming and scattered photons - which become more complicated if the anisotropy is considered, but it is not the purpose of this report to account for them in detail.
However the effect described above can explain the lack of intensity cancellation at a critical angle as can be seen in Fig.~1 and also the fact that the finite residual intensity is excitation energy dependent. \cite{Gozar2}

Although the general effect described here is a standard textbook consequence of light propagation in all anisotropic media, we mention briefly a couple of other examples where it can be important.
NaV$_{2}$O$_{5}$ is an orthorhombic crystal which contains quasi one-dimensional V-O ladders at quarter filling factor. \cite{Grenier}
One of the ways to gain insight in the magnetic and electronic structure might be to look for preferential orientations of dipole moments of infrared excitations with respect to the crystal axes or to perform an angular dependence Raman study \cite{Gozar2} based on similar principles as described in Ref.~\cite{Freitas}.
Potential anisotropy effects which can influence the experimental results should be checked in this case.
Other materials to which such an analysis may also be relevant are the layered Cu-O superconductors with non tetragonal symmetry.
It has been shown for instance that for othorhombic YBa$_{2}$Cu$_{3}$O$_{6+x}$ or Bi$_{2}$Sr$_{2}$CaCu$_{2}$O$_{8+x}$ crystals the real and imaginary parts of the dielectric function are quite different along orthogonal directions in the $ab$ plane \cite{Lance}.
This effect can be relevant for the symmetry analysis of the pair breaking (2$\Delta$) peak below T$_{c}$ and the discussion made for the case of the Sr$_{14}$Cu$_{24}$O$_{41}$ compound with slight modifications can be used to account for the anisotropy in the analysis of the two-magnon scattering spectra in B$_{1g}$ and B$_{2g}$ polarization configurations for YBa$_{2}$Cu$_{3}$O$_{6+x}$ and Bi$_{2}$Sr$_{2}$CaCu$_{2}$O$_{8+x}$ crystals, respectively.
  
I have emphasized the importance of considering the effect of optically induced ellipticity due to the crystal anisotropy when performing polarized Raman or optical spectroscopy.
The discussion is relevant for determining the symmetry and the selection rules associated with elementary excitations.
The few examples referred to in this Comment show that in certain cases this effect should not be overlooked in the data analysis.

I would like to acknowledge useful discussions with G. Blumberg and M. V. Klein.


\begin{references}

\bibitem{Freitas}
P. J. Freitas and R. R. P. Singh, Phys. Rev. B {\bf 62}, 14113 (2000).

\bibitem{Fleury}
P. A. Fleury and R. Loudon, Phys. Rev. {\bf 166}, 514 (1968).

\bibitem{Born}
M. Born and E. Wolf, in \emph{Principles of Optics}, University Press, Cambridge, 1999, p. 790.

\bibitem{Motoyama}
N. Motoyama, and H. Eisaki,  private communications.

\bibitem{Gozar1}
A. Gozar, G. Blumberg, B. S. Dennis, B. S. Shastry, N. Motoyama, H. Eisaki, and S. Uchida, Phys. Rev. Lett. {\bf 87}, 197202 (2001).

\bibitem{Gozar2}
A. Gozar, unpublished.

\bibitem{Grenier}
B. Grenier {\em et al.}, Phys. Rev. Lett. {\bf 86}, 5966 (2001) and references therein.

\bibitem{Lance}
S. L. Cooper {\em et al.}, Phys. Rev. B {\bf 47}, 8233 (1993)

\end{references}
\end{document}